\documentstyle{mn}
\input{psfig}
\begin{document}
\def\ltsima{$\; \buildrel < \over \sim \;$}
\def\simlt{\lower.5ex\hbox{\ltsima}}
\def\gtsima{$\; \buildrel > \over \sim \;$}
\def\simgt{\lower.5ex\hbox{\gtsima}}

\title[The formerly reflection-dominated NGC~6300]
{The formerly X-ray reflection-dominated Seyfert~2 galaxy NGC~6300}

\author[M. Guainazzi]
{Matteo Guainazzi$^1$ \\ ~ \\
$^1$XMM-Newton Science Operation Center, ESA, VILSPA, Apartado 50727, E-28080 Madrid, Spain
}

\maketitle
\begin{abstract}

In this paper, a BeppoSAX observation of the bright Seyfert~2 galaxy
NGC~6300 is presented. The rapidly variable emission from the active
nucleus is seen through
a Compton-thin ($N_H \simeq 3 \times 10^{23}$~cm$^{-2}$)
absorber. A Compton-reflection component with an unusually
high reflection fraction ($R \simeq 4.2$), and the comparison with
a reflection-dominated spectrum measured by RXTE two and half
years earlier suggest that NGC~6300 belongs to the class of
"transient" AGN, undergoing long and repeated periods of low-activity.
The spectral transition provides support to the idea that Compton-thick
and Compton-thin X-ray absorbers in Seyfert~2 galaxies are
decoupled, the former being most likely associated with
the "torus", whereas the latter
is probably located at much larger distances.

\end{abstract}


\section{Introduction}

The X-ray spectra of Seyfert~2 galaxies are almost always seen through
a substantial photoelectric column density
(Awaki et al. 1991; Turner et al. 1997a; see Pappa et al. 2001 for
some recently discovered exceptions). X-ray
absorbers are classified as Compton-thick or -thin, according whether
their column density, $N_H$ is larger than $\sigma_t^{-1} \simeq
1.5 \times 10^{24}$~cm$^{-2}$. Below 10~keV,
the Active Galactic Nucleus (AGN) emission is directly visible in
transmission through a Compton-thin absorber, whereas
it is totally suppressed if the absorber is Compton-thick.
If $N_H \simlt 10^{25}$~cm$^{-2}$, high sensitivity
measurements above 10~keV can still detect the transmitted
emission.

Despite the above classification, the common wisdom so far has
been to associate all the X-ray absorbers with the molecular, dusty
"torus", which encompasses the nuclear environment in the
Seyfert unification scenarios (Antonucci \& Miller 1985; Antonucci 1993).
However, Matt (2000) recently proposed
an extension of the unification theories, where a pc-scale "torus"
is responsible only for the absorption in Compton-thick
objects, whereas the Compton-thin absorber should be located
at much larger scales (10-100~pc), and probably associated with
the host galaxy rather than with the nuclear environment.
The discovery of a plethora of dusty structures in high resolution
HST/WFC2 images of nearby Seyfert galaxies
(Malkan et al. 1998) provides
possible "optical" counterparts to the X-ray absorber{\it s}.

We present in this paper the results of a BeppoSAX
observation of the Seyfert nucleus
hosted in the moderately inclined
($\imath \simeq 51^{\circ}$: Ryder et al. 1996)
barred spiral galaxy NGC~6300 ($z=0.0037$),
where the combination of broadband X-ray coverage and of
a dramatic transition with respect to a Compton-thick,
reflection-dominated state measured by RXTE about two and half
years earlier (Leighly et al. 1999)
provides valuable insights on this issue.

\section{BeppoSAX observation and data reduction}

BeppoSAX observed NGC~6300 from
August 28 (00:41 UTC) to August 29 (16:33 UTC) 1999.
In this paper, only data of the Low
Energy Concentrator System (LECS, 0.5--4~keV;
Parmar et al. 1997), the Medium Energy Concentrator
System (MECS, 1.8-10.5~keV; Boella et al. 1997),
and the Phoswitch Detector System (PDS,
13-200~keV; Frontera et al. 1997) will be presented.
The data were taken from the BeppoSAX
public archive.
Data reduction followed standard procedures,
as described, for instance, in Guainazzi et al.
(1999b). Source scientific products
have been extracted from circular
regions of 8$\arcmin$ and 4$\arcmin$ radii
in the LECS and MECS, respectively. Background
spectra have been extracted from blank sky
fields provided by the BeppoSAX Science
Data Center, using the same region of the
detector as the source. PDS spectra have
been extracted by plain subtraction of the
spectra obtained during the 96~s-long
intervals when
the instrument pointed NGC~6300 and nearby
$\pm$3.5$^{\circ}$ off-axis regions. This ensures a
control of the background systematics
within 0.02 counts per second.
Spectral rebinning ensured
that each spectral bin contains
at least 25 counts and that the instrumental
energy resolution is oversampled by a factor
not higher than 3. Cross-normalization constants
have been added to all the spectral models,
following the prescriptions in Fiore et al.
(1999).
After data
screening, total exposure times were
39.5~ks, 86.2~ks and 77.2~ks for the
LECS, MECS and PDS, respectively. The
corresponding count rates are:
$(1.18 \pm 0.07) \times 10^{-2}$~s$^{-1}$,
$(9.91 \pm 0.11) \times 10^{-2}$~s$^{-1}$,
and $(0.70 \pm 0.04)$~s$^{-1}$, respectively.
The probability that the PDS detection is due
to a serendipitous
contaminating source is $\sim 5 \times 10^{-4}$,
according to the Cagnoni et al. (1998) LogN-LogS.
NGC~6300 ($l = 328^{\circ}$; $b=-14^{\circ}$)
is rather close to the Galactic
Ridge (GR). However, the contribution of the GR
emission to the 10--60~keV
flux in the (1.3$^{\circ}$)$^2$
PDS field of view 
is $\simlt 5 \%$ (Valinia \& Marshall 1998),
and is most likely totally removed in the
background-subtracted spectrum by the
rocking.

In this paper: errors are at 90\% confidence
level for one interesting parameter, and energies
are quoted in the source rest frame, unless
otherwise specified.

\section{BeppoSAX results}

\subsection{Timing analysis}

\begin{figure}
\begin{center}
\psfig{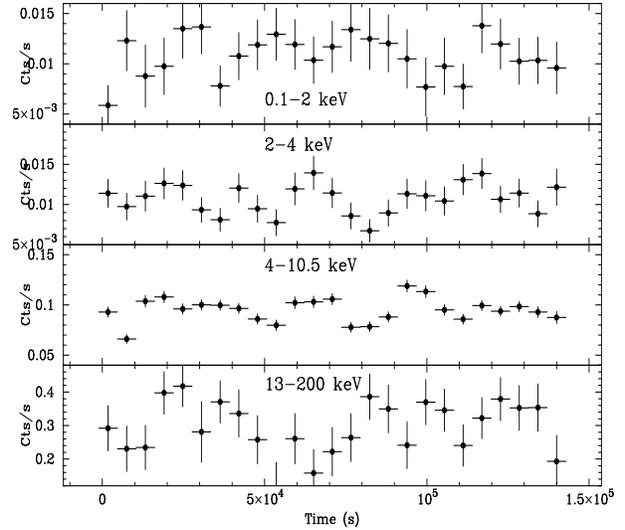}
\end{center}
\vspace{-0.5cm}
\caption{
Light curves of the BeppoSAX observation of NGC~6300 in
the 0.1--2~keV, 2--4~keV, 4--10.5~keV, and 13--200~keV
energy bands (from {\it top} to {\it bottom}). The
binsize is 5760~s. The curves, except the last one, are
not background subtracted.
}
\label{fig1}
\end{figure}

In Fig.~\ref{fig1} we show the light curves
in the 0.1--2~keV (LECS), 2--4~keV, 4--10.5~keV
(MECS) and 13--200~keV (PDS) energy bands.
Each bin ($\Delta t = 5760$~s) corresponds
approximately to one BeppoSAX orbit. Although
some variability is present in
all energy bands above 2~keV, it is
statistically significant only in the 4--10.5~keV
band ($\chi^2_{\nu} = 5.4$ for 25 degrees
of freedom, dof, if a constant line is
fit to the data), where the statistics is
the best.   
%-----------------------------Figure 1--------------------------------
%-----------------------------Figure 1--------------------------------
The X-ray flux in this band is actually
%-----------------------------Figure 2--------------------------------
\begin{figure}
\begin{center}
\psfig{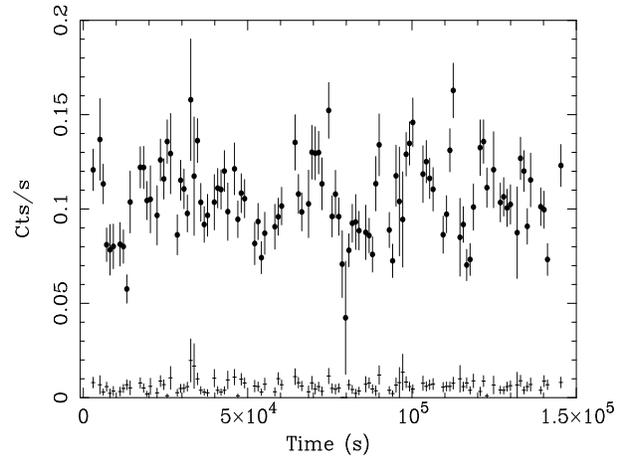}
\end{center}
\vspace{-0.5cm}
\caption{
MECS light curve of the BeppoSAX observation of NGC~6300 in
the 4--10.5~keV energy band. The
binsize is 1024~s. Both the source+background ({\it dots})
and the background light curve (taken from a source-free
region nearby NGC~6300; {\it crosses}) are shown
}
\label{fig2}
\end{figure}
%-----------------------------Figure 2--------------------------------
variable on time scales as low as $\sim$10$^3$~s
(Fig.~\ref{fig2}).
The sharpest variation is a factor of 2 "flare"
of about 2~ks duration, starting
110000 seconds after the beginning of the observation.
No significant spectral variability is nonetheless
observed. Fitting a constant line
to the hardness ratio versus 4--10.5~keV count rate
functions yields $\chi^2_{\nu}=0.85$ and 1.35 for the
4--10.5~keV/2--4~keV and 13--200~keV/4--10.5~keV
bands, respectively.

\subsection{Spectral analysis}

\begin{figure}
\begin{center}
\psfig{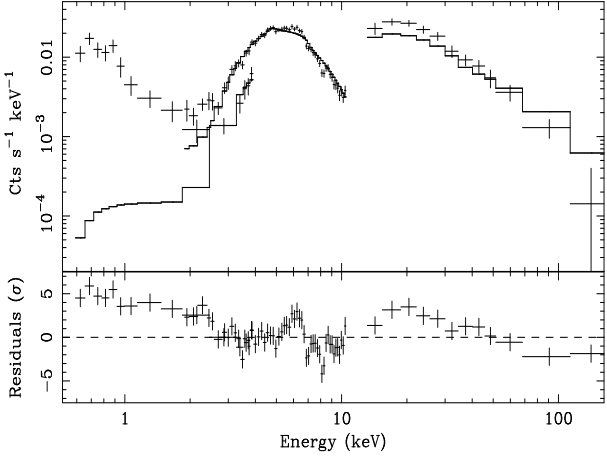}
\end{center}
\vspace{-0.5cm}
\caption{
NGC~6300 BeppoSAX spectra ({\it upper panel}) and residuals in units of
standard deviations ({\it lower panel}) when a
power-law model with photoelectric absorption is
applied.
}
\label{fig3}
\end{figure}

In Fig.~\ref{fig3} we show the BeppoSAX spectrum
%-----------------------------Figure 3--------------------------------
%-----------------------------Figure 3--------------------------------
and the residuals against a power-law fit
with photoelectric absorption. Several
features are evident, and explain the bad fit
($\chi^2 = 385.7/89$~dof). The
MECS spectrum exhibits a low-energy cut-off, due
to a column density
of a few 10$^{23}$~cm$^{-2}$.
This absorbed power-law will be referred to as {\it
Compton-thin power-law} hereinafter.
Both
above 10~keV and below 2~keV excess emission is
present.
The high-energy excess cannot be ascribed to
instrumental effects. The ratio between the
15--200~keV PDS flux and the extrapolation of
the MECS flux in the same energy band is
$1.43 \pm 0.17$, well above the cross-calibration
normalization factor between the two instruments
(0.80-0.90; Fiore et al. 1999).
Additional local features in the 6--8~keV
range suggest the presence of either fluorescent
iron emission lines or photoelectric absorption edges.
A very steep ($\Gamma_{soft} \simeq 4.5$)
power-law absorbed only by the
column density due to our Galaxy ($N_{H,Gal} =
9.4 \times 10^{20}$~cm$^{-2}$; Dickey \& Lockman 1990),
removes the soft excess. In all the
following results we make use
of this model.
Comparably
good fits are, however, obtained if the soft excess is
modeled with an optically thin, collisionally
excited plasma (model {\tt mekal} in {\sc Xspec}),
which may represent the contribution of gas
associated with nuclear starburst emission (Buta 1987),
with $kT \simeq 500$~keV, and
$Z/Z_{\odot} \simeq 5 \%$.
%-----------------------------Table 1--------------------------------
\begin{table*}
\begin{tabular}{lccccccc} \hline \hline
Model & $\Gamma$ & $N^{thin}_H$ & $N^{thick}_H$/R & $E_k$ & $I_c$ & $\Gamma_{soft}$ & $\chi^2$/dof \\ 
& & ($10^{23}$~cm$^{-2}$) &  ($10^{23}$~cm$^{-2}$)/ & (keV) & ($10^{-5}$~cm$^{-2}$~s$^{-1}$) & & \\ \hline
1 & $2.39 \pm^{0.10}_{0.14}$ & $3.1 \pm 0.2$ & $31 \pm^{14}_4$ & ... & ... & $4.4 \pm^{0.5}_{0.4}$ & 110.6/83 \\
2 & $2.32 \pm^{0.12}_{0.13}$ & $2.9 \pm 0.2$  & $30 \pm^8_5$  & $6.27 \pm^{0.13}_{0.16}$ & $4.8 \pm^{1.8}_{1.9}$  & $4.5 \pm^{0.4}_{0.3}$ & 87.8/81 \\
3 & $2.19 \pm^{0.13}_{0.07}$ & $2.10 \pm 0.10$  & $4.2 \pm^{2.6}_{1.7}$  & $6.21 \pm^{0.15}_{0.20}$ & $3.6 \pm 1.7$  & $4.7 \pm^{0.5}_{0.4}$ & 84.3/82 \\ \hline \hline
\end{tabular}
\caption{Best-fit parameters and results for the BeppoSAX observation of
NGC~6300. Explanation of the models is in text}
\label{tab1}
\end{table*}
%-----------------------------Table 1--------------------------------
The quality of the BeppoSAX data does not allow us
to unambiguously characterize the soft X-ray
emission.
The hard X-ray results presented
in this paper are not significantly dependent on
the choice of the soft excess model. 

\begin{figure}
\begin{center}
\psfig{figure=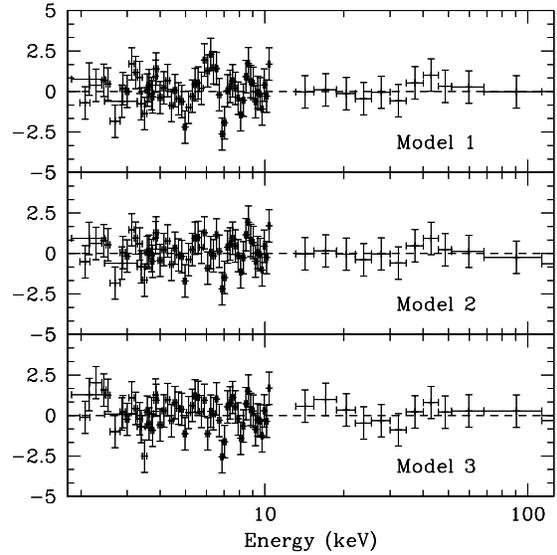,height=8.0cm,width=8.0cm}
\end{center}
\vspace{-0.5cm}
\caption{
2--100~keV residuals in units of standard deviations when
different models are applied to the broadband BeppoSAX
spectra of NGC~6300. The models are explained in text.
}
\label{fig4}
\end{figure}

We have first tentatively interpreted the
high energy excess as the emergence of a
spectral component through a thicker
absorber. We have therefore introduced an additional
absorbed power-law
({\it Compton-thick power-law} hereinafter),
whose index is held fixed
to the Compton-thin power-law.
If the indices are left free in the fit,
indistinguishable best-fit values are yielded. The quality
of the fit is marginally good ($\chi^2 =
110.6/83$~dof; Model~1 in Tab.~\ref{tab1}).
A line-like
feature is still present in the residuals
close to 6~keV (observer's frame; see
Fig.~\ref{fig4}). Adding a
%-----------------------------Figure 4--------------------------------
%-----------------------------Figure 4--------------------------------
narrow ({\it i.e.}: intrinsic width
set equal to 0) Gaussian profile yields a
highly significant improvement in the quality
of the fit ($\Delta \chi^2 = 22.8$ for
two degrees of freedom,
significant at the 99.991\% confidence level;
Model~2 in Tab.~\ref{tab1}). The centroid
energy of the line, $E_k = 6.27\pm^{0.13}_{0.16}$~keV,
is slightly redder than
expected from K$_{\alpha}$ fluorescence
from fully neutral iron (as already noticed
by Leighly et al. 1999),
but consistent within the statistical
uncertainties. The Equivalent Width (EW) is
$140 \pm 50$~eV.
The column density of the thicker absorber
($N_H \simeq 3 \times 10^{24}$~cm$^{-2}$)
justifies our nomenclature. The
cross-section of the thick absorber includes
both the photoelectric and the optically-thin
Compton scattering cross-sections, the latter
playing an important r\^ole at such column densities.
The ratio between
the normalizations
of the Compton-thin and -thick power laws
is significantly smaller than 1:
$N_{thin}/N_{thick} = 2.9 \pm^{1.7}_{1.9} \%$.

The hard X-ray "bump" may be also
alternatively explained in terms of
Compton-reflection of the primary nuclear
emission. We
have therefore substituted the
thick-absorbed power-law with such a component
(model {\tt pexrav} in {\sc Xspec};
Magdziarz \& Zdziarski 1995).
We assume an inclination angle
$\cos \theta = 0.86$. 
The quality of the fit is comparably
good ($\chi^2 = 84.3/82$~dof). The
amount of reflection, $R$
(equal to 1 for reflection of an isotropically
emitted radiation by a plane-parallel,
semi-infinite slab) is, however, unusually
high: $R = 4.2 \pm^{2.6}_{1.7}$.
The exponential cut-off, if left free in
the fit, tends to assume the highest possible
value in the given interval.
The best-fit values in Tab.~\ref{tab1}
correspond therefore
to an ideal case where no cut-off is present.
The 90\% lower limit
on $E_c$ is 250~keV. If $E_c = 250$~keV
(1000~keV),
$R \simeq 2.7$ (3.1) and $\Gamma \simeq 1.99$ (2.09).
The other fit
parameters are only marginally affected by
the change of high-energy model.
The Compton-thin column density is $N_H \simeq
2.1 \times 10^{23}$~cm$^{-2}$, and the
equivalent width of the iron line $100 \pm 50$~eV.

Using Model~3 in Tab.~\ref{tab1}, the observed
fluxes are 0.12, 1.27 and
$9.9 \times 10^{-11}$~erg~cm$^{-2}$~s$^{-1}$
in the 0.5--2, 2--10 and 15--200~keV energy bands,
respectively. They correspond to absorption-corrected
luminosities of 2.4, 1.9 and
$4.9 \times 10^{42}$~erg~s$^{-1}$ in the same
bands, respectively.

\section{Comparison with the RXTE observation}

\begin{figure}
\begin{center}
\psfig{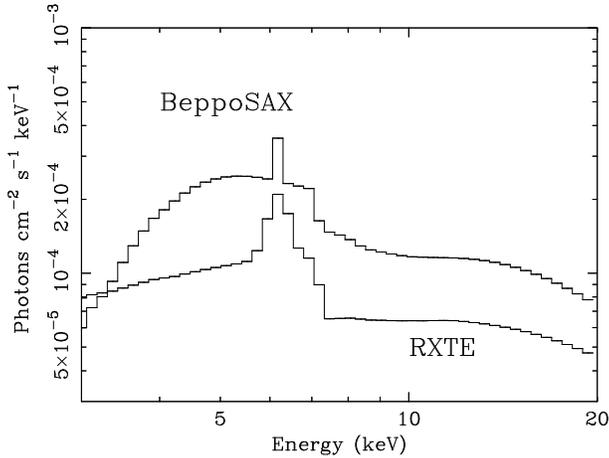}
\end{center}
\vspace{-0.5cm}
\caption{
Best-fit models for the RXTE and BeppoSAX observations
of NGC~6300.
}
\label{fig5}
\end{figure}

In Fig.~\ref{fig5} we compare the best fit models
%-----------------------------Figure 4--------------------------------
%-----------------------------Figure 4--------------------------------
of the RXTE (Leighly et al. 1999) and BeppoSAX
observations of NGC~6300 in the overlapping sensitive
bandpass. As the former we consider
the "Compton-reflection" model in Tab.~1 of
Leighly et al. (1999); as the latter our Model~3
in Tab.~\ref{tab1}. The RXTE observation
caught the source in a significantly
fainter and flatter
state. Together with
the 500-1000~eV K$_{\alpha}$ fluorescence iron
line, this represents a clear signature of a Compton-reflection
dominated spectrum, echo of an otherwise invisible
primary continuum. The BeppoSAX spectrum
is brighter in the {\it whole} 3--20~keV band, {\it
i.e.} in the energy band where {\it both} the Compton-thin
{\it and} the Compton-thick matter play a r\^ole.
Interestingly enough, the intensity of the iron lines
measured by RXTE and BeppoSAX is very close,
and perfectly consistent within the
statistical uncertainties.

\section{Discussion}

In the 2--10~keV energy band, flux
variability on timescales as short as
a few $10^3$~s indicates that the NGC~6300 active nucleus
produces the bulk of the emission in this band.
Indeed, the BeppoSAX spectrum of NGC~6300 in the
same band is
dominated by a standard Seyfert power-law
(the intrinsic spectral index
measured by BeppoSAX is slightly higher, but not extreme,
among Seyfert galaxies; Turner et al. 1997a; Nandra et
al. 1997), seen through a
Compton-thin absorber ($N_H \simeq 2 \times 10^{23}$~cm$^{-2}$).
This model fails, however, to account for the high-energy
part of the broadband BeppoSAX spectrum, where an
additional component, peaking at about
20~keV, is necessary

This component could be explained by the existence of
an optical path to the nucleus, crossing a
Compton-thick absorber ($N_H \simeq 10^{24}$~cm$^{-2}$).
"Dual" or "complex" photoelectric absorbers have already been
observed in intermediate (Zdziarski et al. 2001) or type~2
(Malaguti et al. 1999) Seyferts.
However, this model is not in agreement with the
variability in this source. The difference by a factor 30 between
the normalizations of the Compton-thin and -thick power-laws
in this scenario implies that the absorbing media must
be decoupled\footnote{Actually
this ratio is a only a lower-limit,
because it does not take into account the
contribution of the back-scattering by
the Compton-thick absorber (Matt et al. 1999)}.
The normalization ratio has the correct
order of magnitude, if the Compton-thin power-law
is actually only a fraction of the primary nuclear
emission, scattered
by photoionized plasma (the "warm mirror":
Turner et al. 1997;
Guainazzi et al. 1999a; Awaki et al. 2000;
Sambruna et al. 2001)
along our line of sight.
However, the rapid variability
observed in the 2--10~keV band would require a very
small scattering region: light-crossing arguments would
limit its size to $\simlt 10^{14}$~cm.
The few available variability observations suggest instead
that the "warm mirrors" are located at scales $\sim$1~pc from
the nucleus (Guainazzi et al. 2000), or at least
at distances larger than the typical size
of the Broad Line Regions (BLR; $\simgt 0.01$~pc) if the same
matter is responsible for the broad lines
components emerging
in spectropolarimetric measurements
of narrow-line AGN (Antonucci \& Miller 1985; Tran 1995).
The scattering medium
must extend beyond the Compton-thick matter for the
scattering optical path to be visible.
Last, but not
least, no evidence for iron {\it ionized} K$_{\alpha}$
fluorescence lines is present in the BeppoSAX data.
The EW 90\% upper limit
are 36~eV and 16~eV for He-like and H-like iron, respectively.

A more natural explanation for the BeppoSAX spectrum
is in terms of a Seyfert~1-like spectrum, seen through
the Compton-thin absorber: {\it i.e.}
a Compton-reflection component
superimposed to the primary nuclear emission.
However, the reflection
fraction, a factor of 4 larger than typical
measurements in Seyfert~1s
(Nandra \& Pounds 1994; Perola et al.
in preparation) or Compton-thin Seyfert~2s
(Turner et al. 1997),
cannot be explained without a strong
anisotropy of the nuclear emission or a
delayed response to primary flux changes.

A clue to the origin of
the reflection component may come
from the long-term X-ray history of NGC~6300.
The spectrum observed by BeppoSAX is dramatically different
from that measured by RXTE only two and half years
earlier (Leighly et al. 1999).
The fainter RXTE spectrum is characterized by a very flat
spectral index and a prominent ($EW \simeq$500-1000~eV)
K$_{\alpha}$ fluorescence
iron line. These are clear signatures of a "reflection-dominated"
spectrum, {\it i.e.} a status where the nuclear continuum
is totally disappeared or suppressed, and the bulk of
the hard X-rays is due to the echo of the past nuclear activity,
reflected by Compton-thick matter. In
the standard Seyfert unification scenarios, the reflector
is tentatively associated with the far inner side of the
compact dusty molecular torus, preventing the direct view
of the BLR and of the nucleus in type~2 AGN
(Antonucci \& Miller 1985; Antonucci 1993). If this explanation
is correct (as suggested by Leighly et al. 1999),
the difference with the spectral state measured by BeppoSAX
is due to the later emergence of the nuclear flux.

In this scenario,
there are two possible explanations for the observed spectral
change. It may be due a
Compton-thick absorber with $N_H > 10^{25}$~cm$^{-2}$,
fully covering the nucleus at the time of the RXTE
observation. This 
would have totally suppressed the Compton-thin
power-law. If the Compton-thick matter in the
RXTE observation was reflecting the same average flux
as measured in
the BeppoSAX observation,
a large covering fraction is necessary
to explain the RXTE "echo".
Assuming that 2.5 years represents the timescale
for a single homogeneous $10^{25}$~cm$^{-2}$ cloud to
cross the line-of-sight to the NGC~6300 nucleus, its
distance from the center is
$\sim 0.05 M^{1/3}_7$~pc ($M_7$ is the
black hole mass in units of $10^7 M_{\odot}$). This
would imply a rather peculiar geometry of the Compton-thick
absorber to allow a Compton-thin optical path
to be "freed" during the BeppoSAX observation.

Alternatively, NGC~6300 could represent a case of
"transient" AGN, exhibiting a succession of
high and low activity states. The best monitored
of these transient AGN, NGC4051 (Guainazzi et al. 1998)
remained in a very low state
for about 100~days (Uttley et al. 1999) before returning
its the normal activity. In all other known cases
(Mkn~3, Iwasawa et al. 1994; NGC~2992, Gilli et al. 2000;
NGC~1365, Iyomoto et al. 1997, Risaliti et al. 2001)
the dimming of the X-ray flux is associated with a
dramatic increase of the equivalent width of the
K$_{\alpha}$ iron fluorescent line, suggesting that
the delayed response of a far reprocesser is responsible
for the residual X-ray flux when the primary continuum
is off. In this scenario, the BeppoSAX observation
represents a  snapshot of the dramatic history
in NGC~6300. If the covering fraction of the Compton-thick
matter has not changed, the variation of the flux in
the energy band where reprocessing dominates implies
a factor of 4 difference in the
impinging flux.
Between the RXTE and the BeppoSAX observations,
the AGN switched on again.
However, at the time
of the BeppoSAX observation the active nucleus
was undergoing a new phase of {\it decreasing activity},
not reflected by the Compton-thick matter,
which still echoed a more glorious past.
A dynamical range of at least 10 in
nuclear intrinsic power is necessary to
explain the historical X-ray light curve
of NGC~6300. The real amplitude is likely
to be significant larger, our estimate being
limited by the constraints on
the residual amount of transmitted nuclear flux
still present in the reflection-dominated state.
The Compton reprocessing matter is
constrained by light-travel arguments to be
located within $\simeq 0.75$~pc from the nucleus.
Future campaigns, monitoring the delayed response
of the reprocessed spectral features, may tighten
both the above estimates.

Regardless of which of the above explanations is
correct, the X-ray observations of NGC~6300
have interesting implications on the nature of
the X-ray absorbers in Seyfert~2 galaxies.
Objects like NGC~6300 demonstrate that
Compton-thin and -thick X-ray absorbers may be
physically decoupled. This evidence would
support to the scenarios suggested
by Malkan et al. (1998) and Matt (2000), where the
traditional "torus" is responsible for
the Compton-thick absorbers,
whereas the Compton-thin absorbers should
originate at much larger distances from the
nucleus, maybe associated with the host galaxy rather
than with the nuclear environment. Interestingly
enough, HST high-resolution images of NGC~6300
show an irregular dust distribution in the
innermost 200~pc, with dusty filaments
protruding (and partly covering) the
active nucleus (Malkan et al. 1998).

The above results have also some implications on the
origin of the iron line in Seyfert~2 galaxies. The
intensity of the line in the BeppoSAX observation
(corresponding to an $EW \simeq 100$--150~eV) is in
principle consistent with the line being produced
in transmission by a uniform Compton-thin absorber,
totally encompassing the continuum source
(Leahy \& Creighton 1983). However,
it is also perfectly consistent (within the $\simeq$30\%
statistical uncertainties of each individual
measurement) with the intensity measured in the
RXTE spectrum. This suggests that the bulk of the
iron line observed in Seyfert~2s may be associated
with Compton-thick matter, even when its
continuum spectral signatures are not clearly
visible in the 2--10~keV band, rather than to the
Compton-thin absorber or to the accretion disk.

\vspace{-0.5cm}

\section*{ACKNOWLEDGMENTS}

Fruitful discussions with G.Matt on an early version of this
manuscript are acknowledged.

\end{document}